\documentclass[10pt,conference]{IEEEtran}
\IEEEoverridecommandlockouts

\usepackage{cite}
\usepackage{amsmath,amssymb,amsfonts}
\usepackage{graphicx}
\usepackage{textcomp}
\usepackage{subcaption}
\usepackage{xcolor}
\def\BibTeX{{\rm B\kern-.05em{\sc i\kern-.025em b}\kern-.08em
    T\kern-.1667em\lower.7ex\hbox{E}\kern-.125emX}}
\begin{document}



\title{Generating a Low-code Complete Workflow via Task Decomposition and RAG}
\author{\IEEEauthorblockN{Orlando Marquez Ayala}
\IEEEauthorblockA{\textit{ServiceNow} \\
orlando.marquez@servicenow.com}
\and
\IEEEauthorblockN{Patrice Béchard}
\IEEEauthorblockA{\textit{ServiceNow} \\
patrice.bechard@servicenow.com}
}

\maketitle

\begin{abstract}
AI technologies are moving rapidly from research to production. With the popularity of Foundation Models (FMs) that generate text, images, and video, AI-based systems are increasing their complexity. Compared to traditional AI-based software, systems employing FMs, or GenAI-based systems, are more difficult to design due to their scale and versatility. This makes it necessary to document \textit{best practices}, known as design patterns in software engineering, that can be used across GenAI applications.

Our first contribution is to formalize two techniques, Task Decomposition and Retrieval-Augmented Generation (RAG), as design patterns for GenAI-based systems. We discuss their trade-offs in terms of software quality attributes and comment on alternative approaches. We recommend to AI practitioners to consider these techniques not only from a scientific perspective but also from the standpoint of desired engineering properties such as flexibility, maintainability, safety, and security. 

As a second contribution, we describe our industry experience applying Task Decomposition and RAG to build a complex real-world GenAI application for enterprise users: \textit{Workflow Generation}. The task of generating workflows entails generating a specific plan using data from the system environment, taking as input a user requirement. As these two patterns affect the entire AI development cycle, we explain how they impacted the dataset creation, model training, model evaluation, and deployment phases. 
\end{abstract}


\section{Introduction}
Since the introduction of ChatGPT two years ago, more and more systems based on Generative AI (GenAI) are being shipped to consumers. The underlying technology is a Foundation Model (FM), more commonly a Large Language Model (LLM), able to generate text, images, and, more recently, video.

Due to current limitations in FMs, such as their black-box nature and sensitivity to data distribution shifts \cite{Bommasani2021FoundationModels, zhao2024uncoveringlargelanguagemodel}, building systems leveraging their capabilities is particularly challenging. Integrating AI models into software systems already adds a new set of complex design choices to software engineering \cite{8812912, 10164762, 9734278, 10.1007/978-3-030-19034-7_14, 8804457}. This complexity has increased as the range of possible GenAI-based systems has grown significantly due to their increasing abilities to generate almost any kind of output. For instance, whereas in the past AI models added classification or regression capabilities to software, nowadays systems include more sophisticated features such as code completion, explanation, and search \cite{jiang2024surveylargelanguagemodels}.

The software engineering community has recognized the need for new design patterns for AI-based systems to provide blueprint solutions to the challenges involved in integrating AI models \cite{10164762, 8945075, 10.1145/3487043}. In particular, patterns for architectural design decisions are well-documented \cite{9734278, 10.1007/978-3-031-14179-9_2, 9825909, 10123549}.

Generative AI requires adapting existing design patterns for AI-based systems and adding new ones, as problems specific to generative models such as hallucination (safety) or large number of output tokens (scalability) can affect software systems in new, unexpected ways \cite{10.1145/3663529.3663849}. It is also not trivial to engineer these systems to allow adding features rapidly (modifiability) and to allow them to interact effectively with other non-AI components (interoperability).

As AI practitioners are responsible for building the GenAI models (fine-tuning or prompting), they need to be aware of the impact, positive and negative, that modern techniques have on software engineering \cite{10.1145/3644815.3644957}. However, the typical process to build AI-based systems still sees a rift between scientists and engineers, where the former develop the models and the latter bring the models into production \cite{9734278, 10.36227/techrxiv.21397413.v1}.

The consequence is that models are built with little regard to the desired quality attributes of the overall system, as the main concern of model building is to maximize metrics to prove that the solution works. This may increase the time to deploy the AI-based system into production due to possible redesign of datasets and model rebuilding if the pipeline proves to be too slow, unsafe, etc. Therefore, it is necessary to discuss how common AI techniques affect software engineering.

To help bridge this gap between AI and software engineering, the first part of this paper formalizes two common techniques as design patterns since their utilization impacts the entire engineering effort, crucially the architectural design of the system. Their impact goes beyond constructing the model; hence it is important to discuss how they affect the flexibility, maintainability, safety, and security of the system, among other commonly desired software quality attributes \cite{isoiec250102023}.

Task Decomposition is an application of \textit{divide and conquer} to the AI domain. This technique allows breaking a machine learning (ML) task into sub-tasks, which can be more easily solved. It has been proposed for a variety of tasks such as speech processing \cite{kuan2024speechcopilotleveraginglargelanguage}, film trailer generation \cite{papalampidi2021filmtrailergenerationtask}, and abstract visual reasoning \cite{10.61822/amcs-2024-0022}.

Retrieval-Augmented Generation (RAG) is a well-known technique to allow an FM to interact with data from its environment so that its knowledge is no longer limited by what is stored in its model weights \cite{gao2024retrievalaugmentedgenerationlargelanguage}. RAG has been used with LLMs to reduce hallucination \cite{ayala-bechard-2024-reducing},  to give them access to world knowledge \cite{wang2023knowledgptenhancinglargelanguage}, and to make the generated content more traceable to its sources \cite{xu2024searchinthechaininteractivelyenhancinglarge}.

As our formalization of Task Decomposition and RAG as design patterns derives from our experience in industry, the second part of this paper offers a case study of how we built a GenAI application called \textit{Workflow Generation}. The ML task of generating workflows is similar to generating structured text such as code. A workflow can be seen as a \textit{very specific plan} that automates tasks by interacting with its environment. Since the resulting application was deployed to end-users as part of an enterprise system, it had to meet particular engineering constraints and achieve certain quality attributes.

Using a case study is appropriate as how design patterns are applied depends heavily on the context and constraints in which the system is built. The objective is to show that Task Decomposition and RAG can result in well-engineered systems. We followed traditional guidelines for case studies in software engineering \cite{10.1007/s10664-008-9102-8} to report how these two techniques resulted in good design decisions along the AI development cycle, from data labeling to model deployment. As we are conscious that one can only make limited conclusions from one case study, we invite other AI practitioners to share their work when using these techniques.

Our contributions are the following:
\begin{itemize}
\item We formalize Task Decomposition and RAG as design patterns, documenting and discussing how they impact the engineering of GenAI-based systems.
\item We offer a case study of how we applied these two techniques to build \textit{Workflow Generation}, a GenAI application, to illustrate their positive impact along the entire development cycle.
\end{itemize}

\section{Related Work}
There is a rich body of literature on \textbf{software engineering for ML}, ranging from case studies \cite{8804457, 10.1007/978-3-030-19034-7_14} to surveys drawn from academic and gray literature \cite{10.36227/techrxiv.21397413.v1, 9734278, 10.1145/3487043}. Some papers contribute a list of design patterns for AI-based systems \cite{10164762, 8945075}. A limitation of literature reviews is that they necessarily lack detail, and thus their description of design patterns are at a high-level, divorced from the context in which they are applied. Moreover, most of the existing list of design patterns for ML dates from the pre-generative AI era, when the AI components were relatively simple compared to current uses of FMs. Our work complements this body of work by discussing two techniques in the context of GenAI and providing a detailed description of how we applied them in industry.

\textbf{Foundation Models} are being heavily studied by AI researchers and there is recent work aiming to document how they are currently being inserted in the software stack \cite{10556054}. As FMs are the fundamental AI building block of modern 
GenAI-based systems, there are several architectural design choices to make, such as whether to use an FM hosted internally or hosted by an external provider such as OpenAI \cite{10.1109/MS.2024.3406333}. There are nascent efforts underway to document the new set of challenges FMs add to software engineering \cite{10.1145/3663529.3663849}. Our paper fits into this line of work by discussing how FMs can be used in a real-world application with Task Decomposition and RAG.

\textbf{Task Decomposition} as a technique to solve complex ML tasks is not new. It is related to AI design patterns such as Two-Phase Predictions, AI pipelines, and Multi-Layer Pattern \cite{10164762}. In the GenAI literature, it is related to the Coordination connector, where the FM coordinates tasks between AI and non-AI components \cite{10556054}, and it can be implemented with a Chain of FMs where each FM is responsible for executing a sub-task that is part of a larger task \cite{10.1109/MS.2024.3406333}. We expand on this work by discussing the software quality attributes that this technique offers to GenAI-based systems.

Most modern GenAI systems employ an implementation of \textbf{Retrieval-Augmented Generation}, as a way to improve the output quality of FMs. RAG can be used in an iterative, recursive, or adaptive fashion \cite{gao2024retrievalaugmentedgenerationlargelanguage}, and to adapt FMs to specialized domains \cite{zhang2024raft}. It can be seen as an implementation of the Strategy pattern in the context of AI \cite{10164762}, where the strategy is made up of two interfaces: one to generate text and one to retrieve suggestions used in the generation process. The generation is affected by the quality of the retrieved results, whereas the retrieval is affected by the quality of the query, which can be user-provided or FM-generated. As with Task Decomposition, we formalize its usage as a design pattern and discuss its impact on software quality attributes.

Lastly, automatically \textbf{generating workflows} is an important task for enterprise systems, as building them generally requires expert domain knowledge. Workflows provide significant value to end-users because they automate repeatable processes in several domains such as IT (e.g., acquiring an IT asset) or HR (e.g., onboarding new employees). There exist several approaches to generate them, such as having an LLM generate an initial plan, then asking users to correct the generated plan, followed by asking the LLM again to regenerate the workflow and its details \cite{cai-etal-2024-low-code}. In our use case, we generate a very detailed plan with environment artifacts without any additional input from the user besides the initial requirement. The workflow is generated in a code-like representation that can be rendered in a web user interface.

\section{GenAI Design Patterns}

We consider Task Decomposition and RAG as design patterns because they are reusable solutions to the problem of integrating GenAI components into larger software systems. While they affect all phases of the AI development cycle, they crucially affect the architecture of the system.

Inspired by the template used to document design patterns in the seminal work by Gamma et al. \cite{10.5555/186897}, we describe these two techniques following most of the template's sections, and motivate them using classical software quality attributes \cite{isoiec250102023}.

\subsection{Task Decomposition}\label{TD}
\begin{figure}[htbp]
  \centering
  \begin{subfigure}[b]{\linewidth}
      \includegraphics[width=\linewidth]{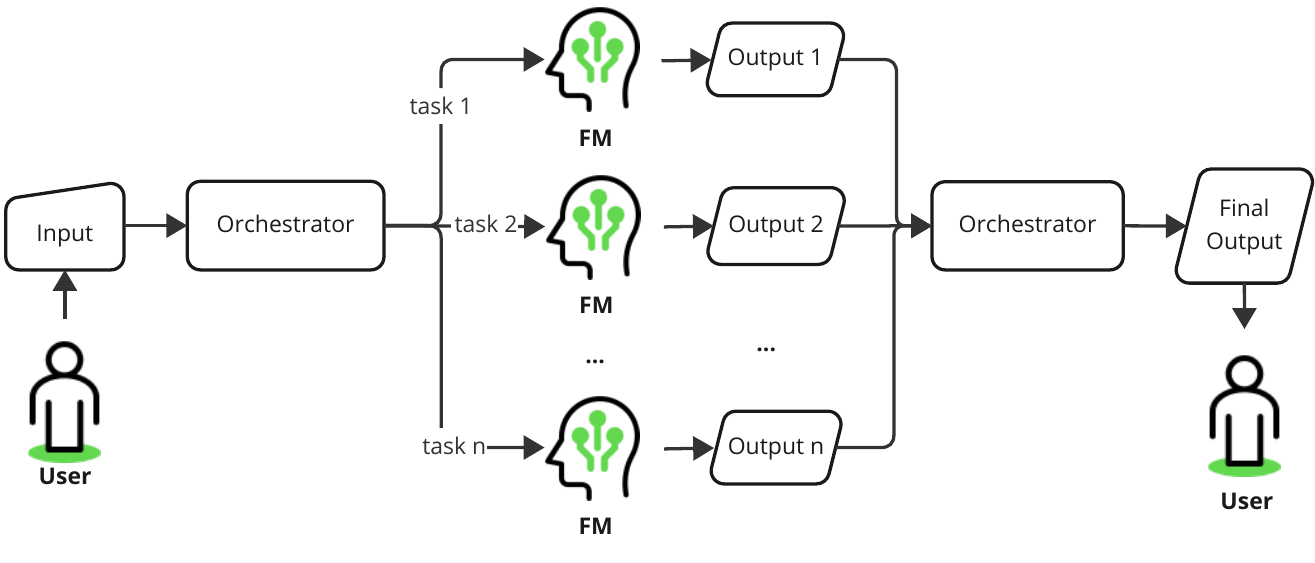}
      \caption{Task is decomposed into sub-tasks that are executed in parallel. Only the final output is shown to the user.}
      \label{fig:td_structure_parallel}
  \end{subfigure}
  \hfill
  \begin{subfigure}[b]{\linewidth}
    \includegraphics[width=\linewidth]{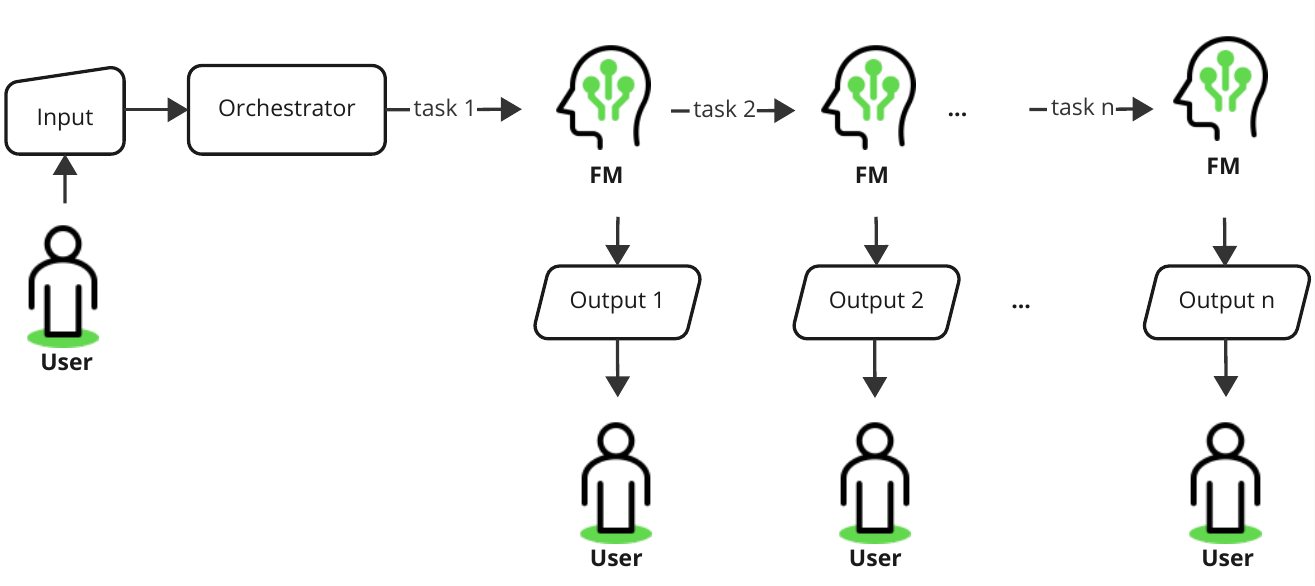}
    \caption{Task is decomposed into sequential sub-tasks whose output are shown to the user, so that users can see the result incrementally.}
    \label{fig:td_structure_sequential}
    \end{subfigure}
\caption{Sample structures of Task Decomposition.}
\label{fig:sample_td}
\end{figure}

\begin{itemize}
    \item \textit{Intent}: Break a complex ML task into simpler sub-tasks.
    \item \textit{Also Known As}: Two-Phase Predictions, AI Pipelines, Multi-Layer Pattern, Chain of FMs.
    \item \textit{Motivation}: 
    \begin{enumerate}
        \item Scalability: FMs can output large number of tokens per input, creating performance risks as the output size increases. Asking the FM to generate output for smaller sub-tasks helps limit scalability issues.
        \item Modularity: By reducing the granularity of the problem, different FMs can solve different parts of the problem, improving the system maintainability.
        \item Functional correctness: As with humans, breaking a problem into sub-problems can aid the FM increase the quality of its output.
        \item Time behaviour: Generating output at the sub-task level reduces the time it takes for end-users to receive a response.
        \item Modifiability: As the model can handle more functionality due to more labeled data or better FMs, more sub-tasks can be added without breaking the overall system design.        
        \item Testability: Evaluating simpler sub-tasks instead of one complex task makes testing FMs easier.
        \item Analysability: It is easier to diagnose problems in the generated output if generation takes place incrementally via sub-tasks.
    \end{enumerate}
    \item \textit{Applicability}: Task Decomposition is suitable for ML tasks that require complex structured output or output that can be structured, such as code generation or essay writing. In the former, when generating a class, the AI model can first generate the signatures of the functions of the class, and then generate the content of each function. In the latter, the AI model can first generate the leading sentence of each paragraph, and then generate each paragraph. In these cases, asking the model to generate the entire output at once may result in lower correctness and/or delays in responses.
    \item \textit{Structure}: Figure \ref{fig:sample_td} shows sample structures.
    \item \textit{Participants}: One or more FMs which execute the sub-tasks, and another component (AI or non-AI) to orchestrate how the sub-tasks are distributed.
    \item \textit{Collaborations}: The orchestrator component divides the sub-tasks among FMs so that the main task is completed. This component can be another FM (e.g., AI agents) or traditional software code (i.e., non-AI component) that calls the FMs deterministically. There also needs to be a mechanism to create the sub-task input, which can come from FMs or non-AI components.
    \item \textit{Consequences}:
    \begin{enumerate}
        \item Decomposition can lead to lower response times and improved system scalability as the size of the FM output is decreased significantly. However, the overall system response time may increase due to the task orchestration.
        \item By having specialized FMs or specialized prompts for the same FM, the overall correctness and modularity of the solution is increased. If there is a sub-task that is more difficult or requires more training/prompting/tuning, then more time and resources can be devoted to it. However, the system complexity increases as there are more components to maintain. For instance, instead of one single prompt, there are several to be maintained and updated. This additional work can be justified only when the ML task is sufficiently complex.
        \item Instead of having to test the overall ML task, test cases or evaluation sets can be created according to the sub-tasks. Error analysis and remediation steps can be conducted according to the different sub-tasks. On the other hand, integration tests are still needed as the output of one sub-task, even if better at the sub-task level, can impact the quality of another sub-task in unforeseen ways.
        \item By making sub-tasks the unit of shippable sub-features, then the system can be adapted to support more and more functionality. However, knowing how to align sub-tasks to sub-features requires deep knowledge of the use case.
    \end{enumerate}
    \item \textit{Implementation}: Decomposing the task may affect data labeling both for training and evaluation. If a dataset is provided at the task level, then it would have to be repurposed so that it can be used for the sub-tasks. Additionally, model deployment can become more complex if more than one FM is used.
    \item \textit{Related Patterns}: The opposite of using Task Decomposition is to let the FM generate the output for the task all at once. This is suitable for ML tasks that are simple or difficult to break into sub-tasks. A pattern related to Task Decomposition is AI Agents, where an FM decomposes the task into sub-tasks and orchestrates their completion (orchestrator agent), and other FMs solve the sub-tasks (worker agents) \cite{tan2024taskgentaskbasedmemoryinfusedagentic}.
\end{itemize}

\subsection{Retrieval-Augmented Generation}\label{RAG}
\begin{figure}[htbp]
  \centering
  \begin{subfigure}[b]{\linewidth}
      \includegraphics[width=\linewidth]{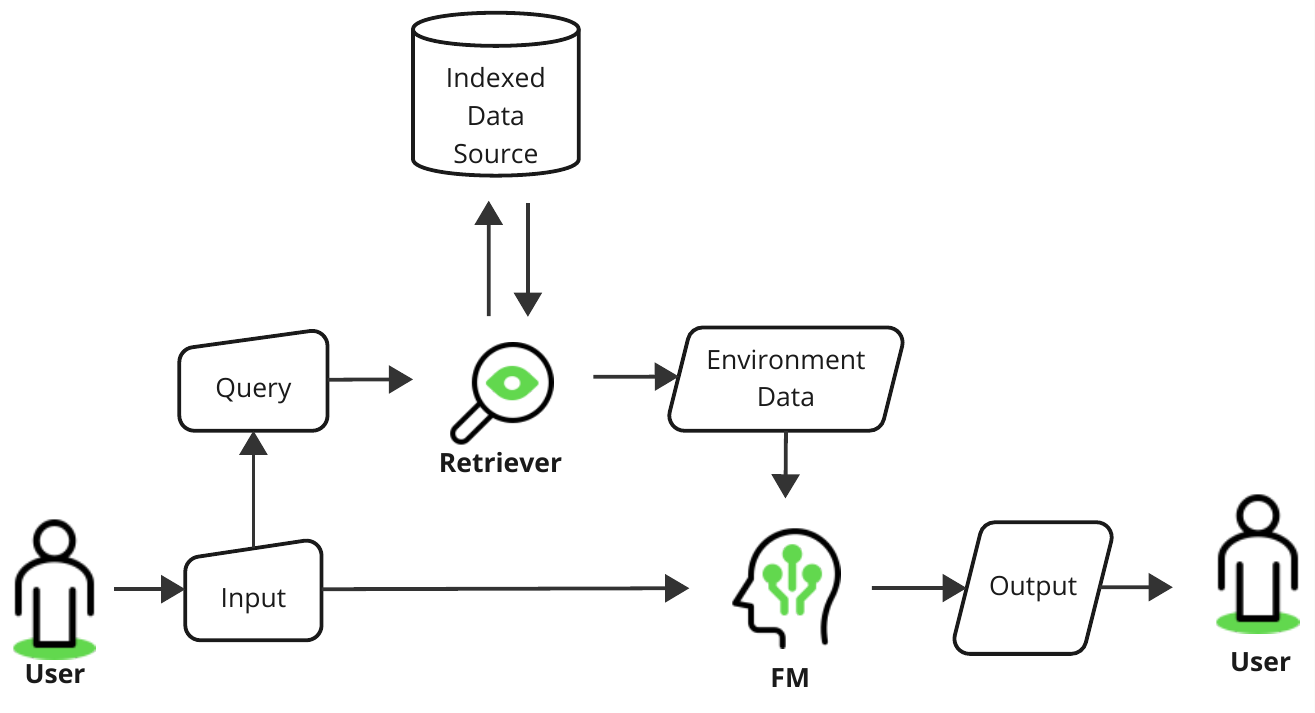}
      \caption{Vanilla RAG where one retrieval takes place per generation.}
  \end{subfigure}
  \hfill
  \begin{subfigure}[b]{\linewidth}
    \includegraphics[width=\linewidth]{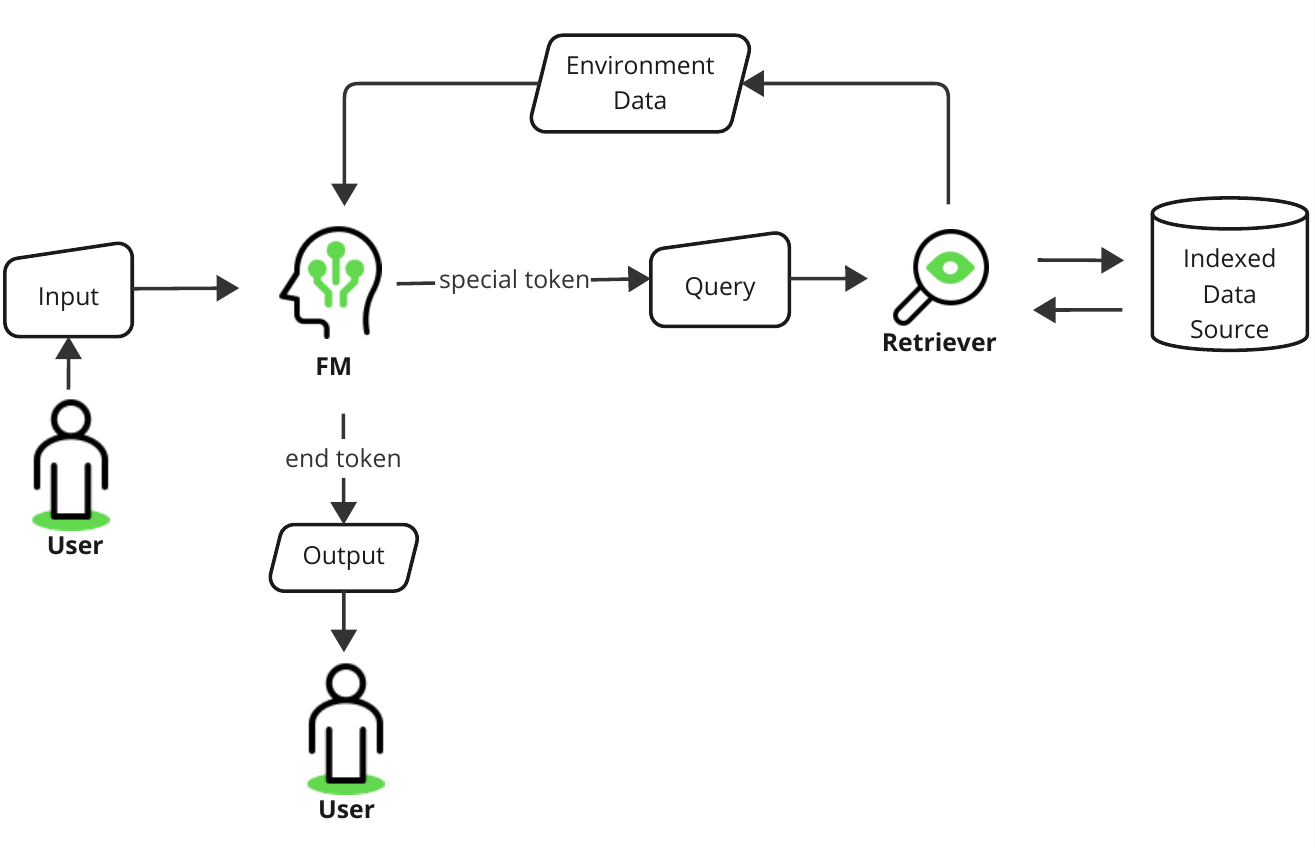}
    \caption{FM decides when to request for environment data based on outputting special tokens, until an end token is reached; also known as adaptive RAG \cite{gao2024retrievalaugmentedgenerationlargelanguage}.}
    \label{fig:rag_structure_adaptive}
    \end{subfigure}
\caption{Sample structures of Retrieval-Augmented Generation.}
\label{fig:sample_rag}
\end{figure}

\begin{itemize}
\item \textit{Intent}: Provide environment data to an FM so that it can adjust its output according to real and available input.
    \item \textit{Also Known As}: RAG.
    \item \textit{Motivation}: 
    \begin{enumerate}
        \item Security: FMs can introduce security risks if allowed to call functions or tools available in the environment \cite{cohen2024jailbrokengenaimodelcause, wu2024darkfunctioncallingpathways}. By suggesting environment data in the input to the FMs, they can interact with the environment in a more secure way.
        \item Safety: Retrieving and suggesting facts to an FM during generation can decrease the amount of hallucination in their output.
        \item Modularity: To simplify the ML task for the FM, it is desirable to separate \textit{how} the task is solved from \textit{what} data is needed to solve it. Requiring the FM to memorize all the data it needs for the task in its model weights is unfeasible.
        \item Interoperability: Modern AI systems require FMs to interact with artifacts or knowledge available in the environment in which they run.
        \item Functional correctness: The knowledge embedded in the model weights of an FM is limited to the data it has received during training. By augmenting its knowledge with information from the environment, it can increase its correctness.
        \item Self-descriptiveness: GenAI-based systems tend to be black boxes; showing the retrieval results gives users more visibility on the system's inner workings.
        \item Testability: RAG allows evaluating the retriever and the indexed data sources independently of the FM.
        \item Analysability: When errors arise, RAG allows a diagnostic of whether the FM made an error due to lack of AI capabilities, or due to inappropriate environment data available to it.
    \end{enumerate}
\item \textit{Applicability}: RAG is useful when an FM needs to access up-to-date knowledge from its environment, allowing it to generalize to scenarios or input data not seen during its training. When FMs are not grounded to the environment in which they run, they have a higher likelihood of producing answers that are plausible yet false. When models tend to hallucinate, it is a good idea to use RAG.
\item \textit{Structure}: Figure \ref{fig:sample_rag} shows sample structures.
\item \textit{Participants}:
    \begin{enumerate}
        \item A retriever, normally another AI model, used to encode the data to be retrieved as well as the search query. A simpler keyword search mechanism can be used to reduce the complexity of the solution \cite{10.5555/188490.188561}.
        \item An FM which will receive data from the retriever in its input prompt. This FM can also be used to create the queries that will be used to retrieve data from the environment.
        \item Sources of data where the retriever pulls data from: databases, text documents, the web, etc.
\end{enumerate}
\item \textit{Collaborations}: There needs to be a mechanism to index the environment data so that it can be efficiently retrieved. The retriever is used to create embeddings, or vectors, that can be indexed offline. At generation time, when retrieval needs to happen, this same retriever will create an embedding for the query. A similarity (e.g. cosine similarity) score will be created to retrieve the indexed data that is most relevant to the query. This data is then passed to the FM during generation. There are several ways to orchestrate this collaboration between retrievers and generators: iterative, recursive, and adaptive \cite{gao2024retrievalaugmentedgenerationlargelanguage}.
\item \textit{Consequences}:
    \begin{enumerate}
        \item As the FM will receive suggestions of what data it should include in its output, the likelihood of hallucinated and unsafe output decreases. However, this requires the sources of data to be kept up-to-date and free of unwanted output, or having another mechanism to filter out the suggestions.
        \item The responsibilities for providing up-to-date environment data to the FM will rest on the retriever and the indexed sources of data, allowing the FM to be optimized for solving the ML task. This separation of concerns increases modularity and makes it clear that only the retriever AI component interacts with sensitive non-AI components, such as databases. The trade-off is higher system complexity as now there are two AI components to maintain: the retriever and the FM.
        \item The system is likely to be more correct, especially if its functionality depends on access to latest data or knowledge. However, there is no guarantee that the retriever suggestions are of good quality or that the FM will accept them during generation \cite{10.1145/3644815.3644945}. Possible approaches to ameliorate this include offering many suggestions to the FM instead of one, training the FM to receive suggestions in a particular form, and jointly training the retriever and the FM.
        \item Showing the retrieval suggestions to end-users along with the generated output allows them to decide whether the output is correct. This provides a measure of self-descriptiveness as the generation process is more transparent, especially in knowledge-based ML tasks such as summarization and question-answering. However, showing why the FM used or ignored the suggestions is currently not feasible, as explaining the output of an FM is still an open research question.
        \item The separation of concerns between retrieval and generation allows testing each of these AI components separately. Retrieval is generally evaluated with metrics such as Recall, Hit Rate or Mean Reciprocal Rank \cite{gao2024retrievalaugmentedgenerationlargelanguage} whereas the metrics used to test the overall system output resulting from generation depend largely on the use case.
        \item Adding a retriever in the generation process not only adds one more AI component to maintain but also additional latency, which can detrimentally impact the time behaviour of the system.
    \end{enumerate}
\item \textit{Implementation}: The design of the retriever is as important as the design of the FM. For domain-specific use cases, it may be required to train a retriever model on domain-specific data to achieve good suggestions \cite{zhang2024raft}. If latency is a concern, simple retrievers based on term-frequency can be used \cite{10.5555/188490.188561}. When setting up the generation phase using the FM, special care needs to be taken to allow the generation to continue, albeit at lower quality, even if the retriever step fails or the suggestions are of little use. While jointly training the retriever and FM can increase output quality \cite{wang2024jmlrjointmedicalllm}, this increases the coupling between both AI components, as the retriever may perform poorly when another FM is used.
\item \textit{Related Patterns}: Tool Usage is another approach to allow FMs to access environment data \cite{10.5555/3666122.3669119}. FMs are trained to make API function calls to access external tools such as search engines and calculators. This approach is more flexible as the FM can call any API available in the environment yet it introduces more security risks as arguments passed to these function calls may result in unauthorized output \cite{cohen2024jailbrokengenaimodelcause, wu2024darkfunctioncallingpathways}. A combination of RAG and Tool Usage would be to suggest APIs and API arguments to the FM, but this does not guarantee that the FM will restrict itself to using the suggested arguments. Tool Usage, however, removes the need to have another AI retriever component and maintain indexes of environment data. Another much simpler approach is to simply include the environment data in the prompt, but this is only feasible when this data is static or limited; otherwise the FM prompt can become too large. Fine-tuning the FM can help it memorize environment data but the FM would need to be fine-tuned every time the environment data changes significantly \cite{10.1145/3644815.3644945}.
\end{itemize}

\section{Workflow Generation: A Case Study}
Having discussed Task Decomposition and RAG at length, we report how our team built a real-world GenAI application called \textit{Workflow Generation}. We structure our reporting as a descriptive case study \cite{10.1007/s10664-008-9102-8}, where the objective is to show how these design patterns result in a well-engineered AI system.

The context of this case study is a Platform-as-a-Service enterprise company that automates the creation and execution of workflows in domains such as IT and HR. The core team responsible for the GenAI application is made up of a dozen individuals, including software engineers, data labelers, quality engineers, and AI applied researchers. The application is shipped as an API that is served in the enterprise cloud platform. While another team is responsible for building the user interface, we restrict the case study to only the core team.

There are two research questions that we attempt to answer:
\begin{enumerate}
    \item[RQ1.] \textit{How does Task Decomposition and RAG impact data labeling, model training, model evaluation, and model deployment?}
    \item[RQ2.] \textit{What are the trade-offs involved when using these two techniques?}
\end{enumerate}

L.E. Lwakatare et al. \cite{10.1007/978-3-030-19034-7_14} grouped the software engineering challenges common to commercial AI-based systems in four categories: Assemble dataset, create model, train and evaluate model, and deploy model. Hence, RQ1 is meant to address how these techniques help deal with these challenges in the GenAI era, in terms of software quality attributes.

RQ2 aims to elicit discussion on limitations of these techniques as well as on alternative approaches that were not chosen to build the application. It is easier to understand trade-offs when real-world constraints and requirements are considered.

The sources of information for this case study are mainly of first degree, as the authors of this paper were key contributors in the entire development cycle. Third degree sources, such as datasets, design documents and experiment journals, complement the self-reporting.

Before answering both research questions, we will define the ML task.

\subsection{ML Task Definition}
A workflow automates a repeatable process by listing the steps that need to be executed in a given order. Each of these steps translates to code, but end-users avoid writing code by building a workflow; this is referred to as \textit{low-code}.

Each environment, or installation of the enterprise platform, defines a set of steps $S=\{s_1, s_2, ..., s_n\}$. Each step $s$ has a set of inputs $I_s=\{i_1, i_2, ..., i_n\}$ and a set of outputs $O_s=\{o_1, o_2, ..., o_n\}$. Inputs and outputs have a \texttt{name} and a \texttt{value}, but the \texttt{name} of an input as well as the \texttt{name} and \texttt{value} of outputs are deterministic (i.e., they do not need to be generated by the FM). Only the \texttt{value} of inputs change depending on what the workflow does.

A workflow $w$ is an ordered list of steps $S_w=(w_1, w_2, ..., w_n)$ where every element in $S_w$ comes from set $S$. In $w$, the inputs of $w_n$ can use the outputs of $w_m$ where $m<n$, that is, the outputs of previous steps can be used in future steps. Importantly, the input values of $w_n$ can also use environment artifacts such as database table names, table columns and row values.

The ML task then consists of generating workflow $w$ from a natural language text such as \textit{When a P1 incident is created, look up the user assigned to the incident and if the user has a manager, send an email reminding them of the incident}.

\begin{figure}[htbp]
  \centering
  \includegraphics[width=\linewidth]{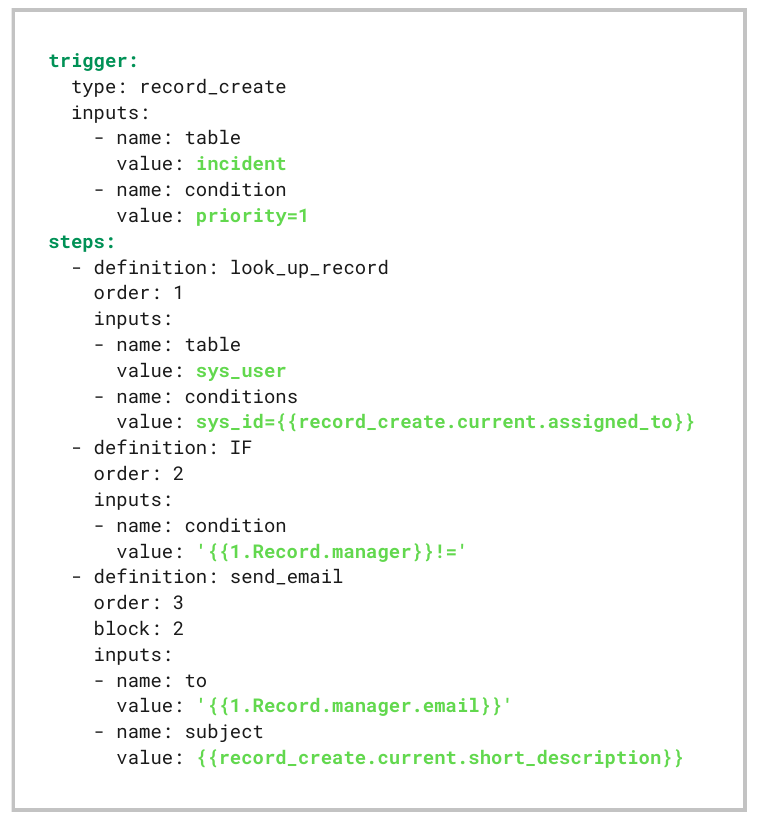}
\caption{Sample workflow mapping to the user requirement \textit{When a P1 incident is created, look up the user assigned to the incident and if the user has a manager, send an email reminding them of the incident}. Values coming from the system environment are in light green (prefixed by key \texttt{value}).}
\label{fig:sample_task_example}
\end{figure}

Figure \ref{fig:sample_task_example} shows the expected workflow in YAML format. This representation took several iterations of discussion among team members, as we needed readable workflows for labeling and to make the ML task as intuitive as possible without requiring a large number of tokens. There are four steps:
\begin{enumerate}
    \item The trigger step determines when the workflow should execute. In this example, it executes when a record in the \texttt{incident} database table is created but only if the \texttt{priority} value in the record is \texttt{1}.
    \item The second step has \texttt{order: 1} (the trigger step can only be in the beginning of the workflow; thus it has no \texttt{order} value). This step reads a record from the \texttt{sys\_user} table given a condition, which uses an output of the trigger step. Step outputs are represented by having their names inside \texttt{\{\{} and \texttt{\}\}}.
    \item The step at \texttt{order: 2} tests for a condition using an output from step at \texttt{order: 1}.
    \item The last step sends an email only if the condition of the previous step is true, as its \texttt{block} value is 2, denoting that it is a child of the \texttt{IF} step, which has \texttt{order: 2}
\end{enumerate}

During our first round of prototyping, we determined that this ML task is very difficult, even with GPT-3.5, a state-of-the-art FM at that point in time. The main challenges are:
\begin{itemize}
\item Mapping ambiguous text to specific steps: each serves a particular function and has a unique set of possible inputs.
\item Being aware of the workflow's building blocks that are specific to the system environment: steps, database tables, columns, values, etc.
\item Understanding the steps' interdependencies: step $N$ can use outputs coming from the previous $N-1$ steps.
\item Generating a complete workflow with acceptable latency: they can be long in number of steps (more than 20), and thus require a large number of tokens.
\end{itemize}

In order to ship a first version of the application, we had to break the task into two sub-tasks:
\begin{enumerate}
    \item \textbf{createFlow}: Generate the outline of the workflow, listing only the step names and their order/block in the list.
    \item \textbf{populateInputs}: Generate the inputs of a single step given previous steps.
\end{enumerate}
We then realized that we needed RAG to achieve acceptable quality as the rate of hallucination of step names and database table names was more than 20\% in our first experiments. The following subsections will detail how these two decisions to use Task Decomposition and RAG resulted in shippable, well-engineered software.

\subsection{Data Labeling}
While labeling is expensive, it is necessary for achieving automatic evaluation. When prompting does not yield good results, the last resort is to fine-tune an FM with labeled training data. Part of the team tried to write effective prompts, an acknowledged challenge in the software engineering community \cite{10.1145/3663529.3663849}. The conclusion was that in-context learning via a long description and few-shot samples only worked for simple workflows. We thus decided to fine-tune an FM.

We had thousands of internal workflows that we could use for labeling, to train and evaluate the model. As this is a domain-specific use case, we had to write clear labeling guidelines and train data labelers. To ship an initial version of the GenAI application, we needed to have labeled samples fast. However, training labelers to write a text description (requirement) for a complete workflow such as the one shown in Figure \ref{fig:sample_task_example} would have taken too long due to the large amount of information present in workflows.

How the task is framed affects heavily the labeling effort. This is where Task Decomposition is useful. When the use case permits it, it is best to break down the labeling effort into multiple rounds where one round maps to a releasable version of the application. In our case, we decided to first ship \textit{Workflow Generation} only with the \textbf{createFlow} sub-task.

\begin{figure}[htbp]
  \centering
      \includegraphics[width=\linewidth]{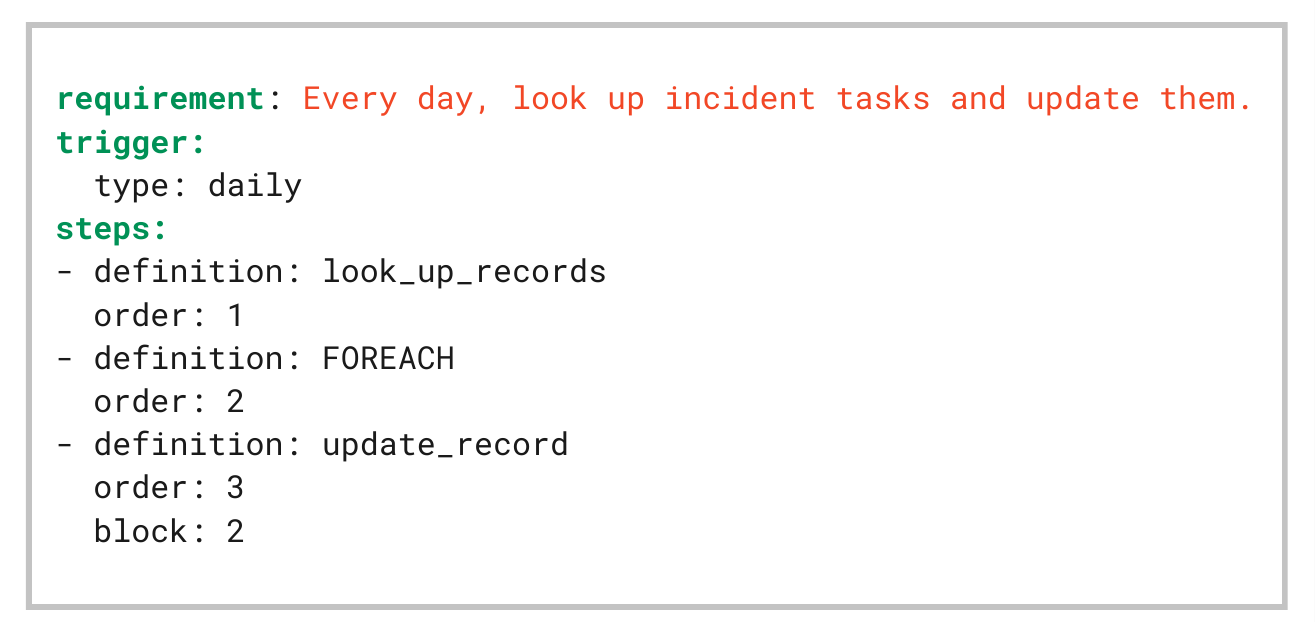}
      \caption{Labeling \textbf{createFlow} samples involved writing the requirement (in red) given the basic workflow outline.}
      \label{fig:create_flow_subtask}
\end{figure}

Our chosen decomposition allowed us to create a dataset for the first task fast so that we could start validating whether a fine-tuned FM could at least generate the outline. Figure \ref{fig:create_flow_subtask} shows a sample for the \textbf{createFlow} sub-task. Compared to generating the entire workflow, there are two advantages. First, it is much simpler to label as the labeler only had to write a requirement for the steps, ignoring the inputs. Second, it is an easier task for the FM as only the list of steps is generated.

\begin{figure}[htbp]
  \centering
    \includegraphics[width=\linewidth]{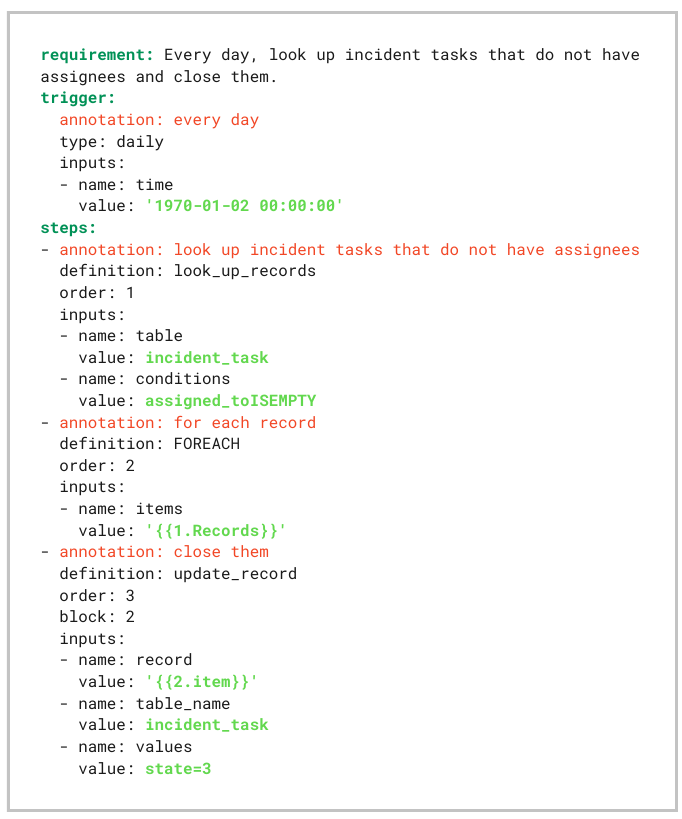}
    \caption{Labeling \textbf{populateInputs} samples involved writing one \texttt{annotation} (in red) per step where the annotation contains enough detail to generate the \texttt{inputs} values (in light green).}
    \label{fig:populate_inputs_subtask}
\end{figure}

Figure \ref{fig:populate_inputs_subtask} shows the more complex labeling task for \textbf{populateInputs}. In this case, the labeler had to write a label for each step, as an \texttt{annotation}, which is normally text that is part of the \texttt{requirement}. This was more time-consuming because the labeler needed to understand the values in the step inputs, including column names, column values, and operators. The complexity can be seen even in the simple examples in Figures \ref{fig:create_flow_subtask} and \ref{fig:populate_inputs_subtask}. The \texttt{requirement} evolved from \textit{Every day, look up incident tasks and update them} to \textit{Every day, look up incident tasks that do not have assignees and close them}. The labeler had to map \texttt{assigned\_toISEMPTY} to \textit{do not have any assignees} and \texttt{state=3} to \textit{close them}.

While at the end labelers had to master the intricacies of workflows, Task Decomposition helped phase this learning into two parts, thereby allowing us to go to market fast and incrementally improve the shipped solution. A trade-off in labeling is that dataset management becomes more complex, as now there are several datasets to verify and maintain.

Task Decomposition also contributed to the modifiability of the system. To continue enhancing the application iteratively, we first asked for labels for the most common steps, as these are simpler to understand. Supporting more steps became simply a matter of labeling less common steps, once the labelers became more experienced.

We did not need any additional labeling for RAG. Data to train the retriever model was obtained by leveraging the same data. The \textbf{createFlow} dataset was used to create a RAG dataset for retrieving step names given a user requirement, whereas the \textbf{populateInputs} dataset was used for retrieving table names, column names, and column values given an annotation.

\subsection{Model Training}
The greatest impact of Task Decomposition and RAG is on model training, or model fine-tuning in our case, as how models are trained determine how they will interface with the rest of the system (i.e., model inputs and outputs). Using samples such as the one shown in Figure \ref{fig:create_flow_subtask}, we shipped a first version of the application that could generate only outlines, with low rates of hallucination thanks to a simple RAG implementation. But the real challenge was to generate the complete workflow.

The first step was to find a good enough retriever. As our use case is domain-specific, available open-source retriever models, such as GTR-T5 \cite{ni-etal-2022-large}, did not provide satisfactory results. For scalability and latency reasons, we chose a small retriever of about 100 million parameters that encoded text in tens of milliseconds. Using our labeled data, we fine-tuned this small retriever via a common technique called contrastive learning \cite{1640964, gao-etal-2021-simcse}, thereby obtaining good retrieval quality.

The next step was to fine-tune the FM to perform the sub-tasks. Given that we were not prompting an off-the-shelf FM, we had flexibility on how to train the FM to do RAG. To decrease latency, it is desirable to minimize the number of retrievals. This is accomplished with a method known as adaptive retrieval \cite{gao2024retrievalaugmentedgenerationlargelanguage}, where the FM requests additional data, when needed, to continue the generation (Figure \ref{fig:rag_structure_adaptive}). This can result in higher output quality as the FM is trained to rely on the data it has memorized during training for input that is easy or very frequent. For input that is more complex or infrequent in the data distribution, the FM relies on RAG.

\begin{figure}[htbp]
  \centering
    \includegraphics[width=\linewidth]{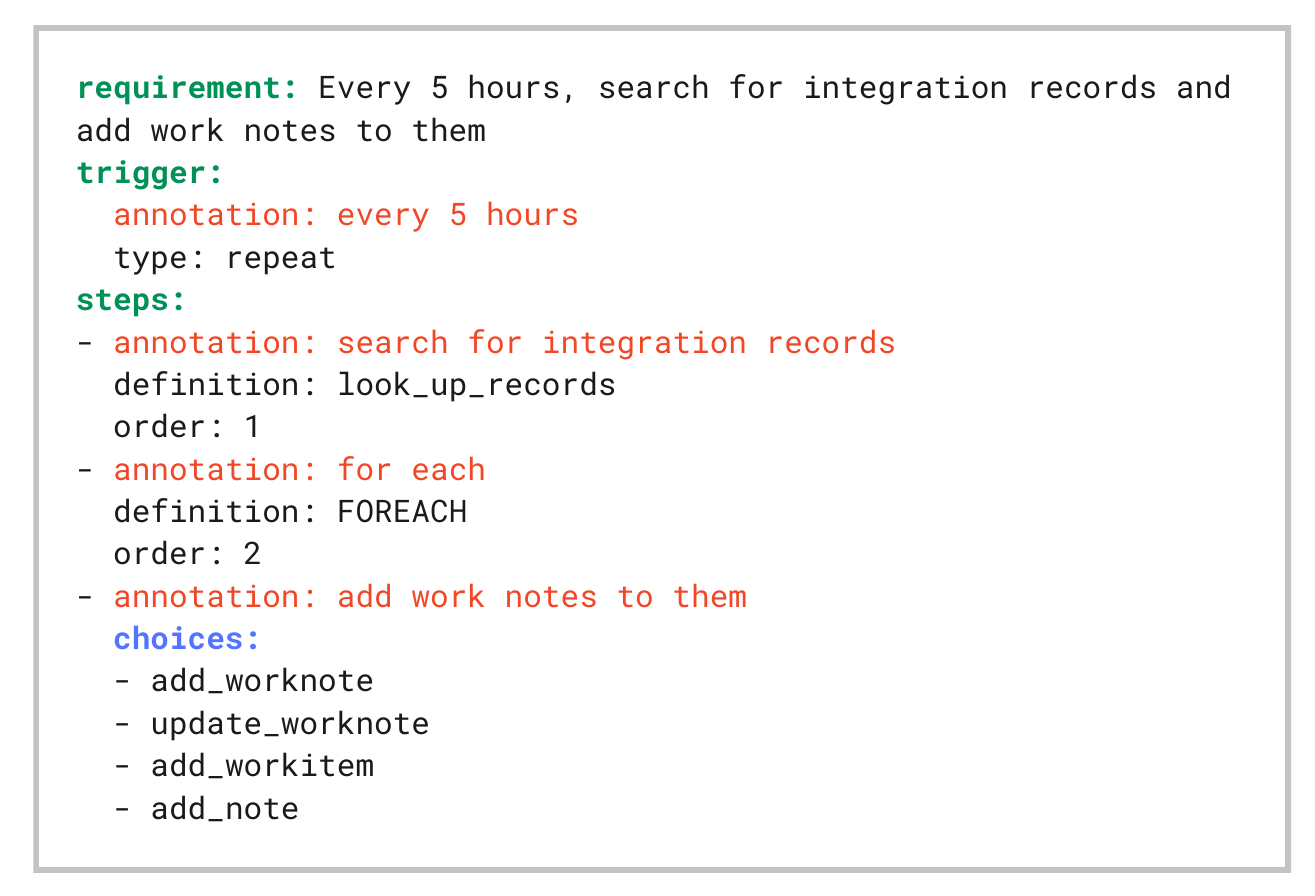}
    \caption{Example of how RAG was used in the training process. Text \textit{choices:} in blue is a special token that the FM uses to signal that it wants suggestions from the environment. Here, the retriever module offers four choices for the step name.}
    \label{fig:rag_choices}
\end{figure}

Figure \ref{fig:rag_choices} shows an example of how the labeled samples for \textbf{createFlow} were transformed to suggest choices to the FM coming from the environment. We used \texttt{choices:} as a special token to stop generation and call the retriever using the step annotation, in this case: \textit{add work notes to them}. At inference time, when this token is generated, the retriever module is invoked. The samples for \textbf{populateInputs} were similarly transformed to give choices to the FM for table names, column names, and column values. Note that, akin to teacher forcing \cite{6795228}, during training the FM will always see the ground-truth value in the list of choices.

Another important design choice when using Task Decomposition is how to configure the link between the sub-tasks. A crucial difference between the example in Figure \ref{fig:create_flow_subtask} and Figure \ref{fig:rag_choices}, besides \textit{choices:}, is that the latter has a field called \textit{annotation}, which describes the steps. In a later version of the \textbf{createFlow} dataset, the FM is fine-tuned to generate these annotations by extracting them from the user requirement. These annotations then become a crucial part of the \textbf{populateInputs} training samples, as they are used to populate the environment artifacts in the step inputs.

To orchestrate the sub-tasks, we followed an approach similar to Figure \ref{fig:td_structure_sequential}, where the orchestrator is a simple non-AI component that first calls the \textbf{createFlow} API. Then it loops over all steps in the outline, generating the inputs by calling the \textbf{populateInputs} API for every step. Via this orchestration, the complete flow is generated.

Using Task Decomposition and RAG resulted in three modules: one to generate the outline, one to populate the step inputs, and one to retrieve suggestions. To simplify the system, we use the same fine-tuned FM for both sub-tasks, but the benefits in modularity would allow us to use a larger FM for the more complex problem of generating the outline and a smaller FM for the other task. This would help the scalability of the system because smaller FMs have lower latency and there are many more API calls to populate inputs.

This modularity also gives the system more modifiability and replaceability. We can improve the AI capability to generate the outline independent of the rest of the system and we can retrain the module that populates inputs, once we have more labeling data, so that it supports the less frequent steps.

Decomposing the tasks also helped the time behaviour of the system, as the user can see the outline before deciding to continue the generation. The interaction of the system is further enhanced by allowing the user to stop generation and modify their requirement if the generated outline is erroneous.

A measure of self-descriptiveness, or interpretability, is achieved with the generated annotations, but especially with RAG, as we could show users the choices that were considered by the FM.

Besides the reusability that RAG provides, since the retriever model was fine-tuned for all environment artifacts, RAG improves the security and safety of the system. By using RAG, we ensured that the FM does not contain any customer-specific knowledge in its model weights. The retriever choices can be filtered according to user permissions (confidentiality) and the FM can only use data that the retriever suggests (integrity). Lastly, users can judge the risk of the generated output by consulting the retriever choices (risk identification).

The disadvantages of Task Decomposition in model training is that it becomes multi-task training, which adds complexity such as deciding the proportion of samples for each task and can make training more unstable. Normally a sufficiently large dataset can alleviate these issues. We also had to fine-tune our own retriever model, which required more effort and time. This extra fine-tuning effort can be justified only by the complexity of the workflow generation task.

\begin{figure}[htbp]
  \centering
    \includegraphics[width=\linewidth]{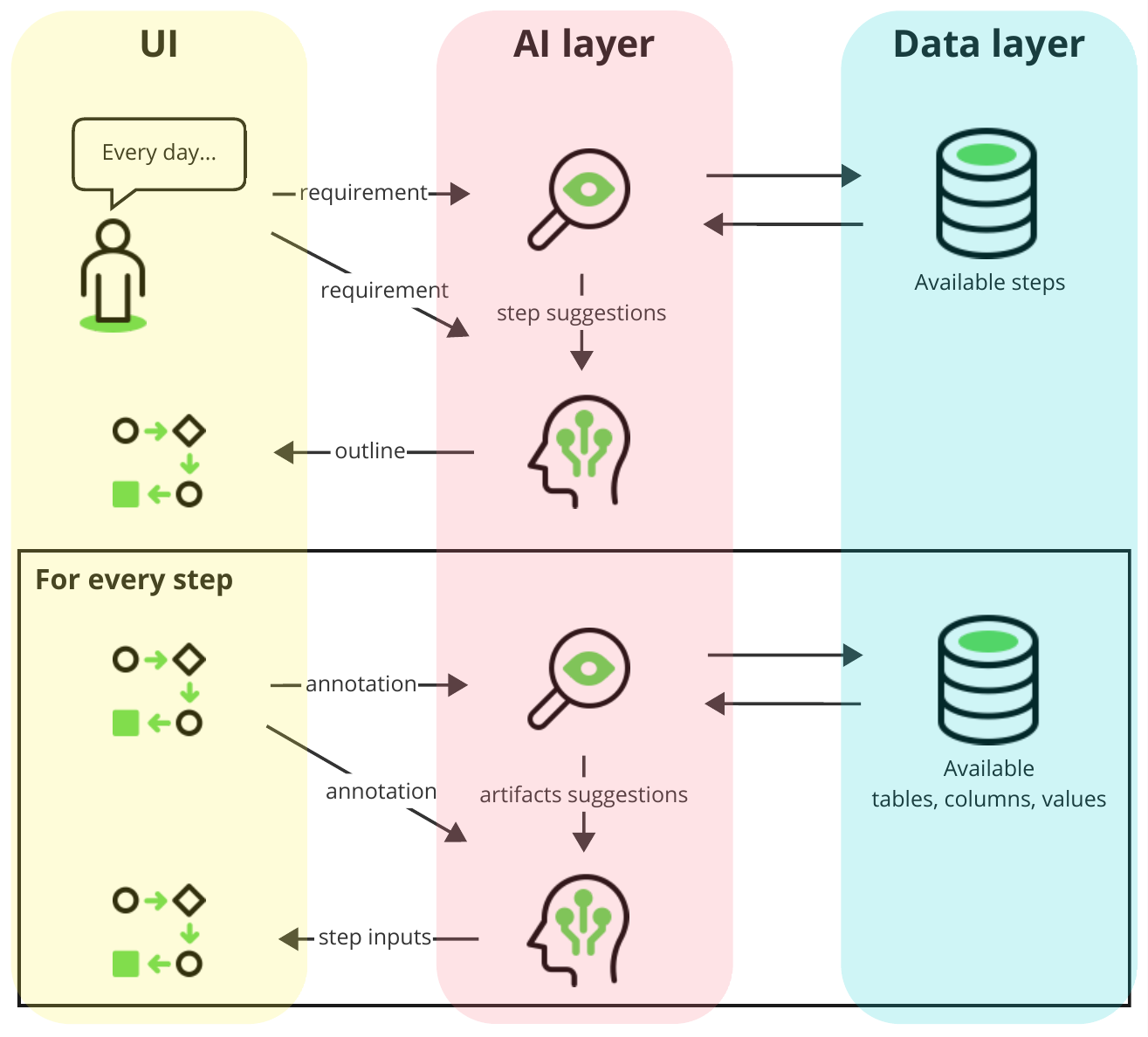}
    \caption{System architecture with UI, AI, and data layers.}
    \label{fig:architecture_diagram}
\end{figure}

Figure \ref{fig:architecture_diagram} shows the three-layer architecture of the system, resulting from decisions made in the model training phase. The UI receives the user requirement and displays the workflow outline and step inputs as they are populated. The AI layer contains the retriever and the fine-tuned FM. The data layer stores the indexed sources of data, from where the retriever suggests choices to the FM. This layer can be replaced in every installation of the platform to allow the FM to generate output specific to each customer, without compromising security.

\subsection{Model Evaluation}
Evaluating the model is as important as data collection. It is generally acknowledged that testing GenAI-based systems is particularly challenging due to the open nature of generation \cite{abeysinghe2024challengesevaluatingllmapplications}. There is generally more than one generated answer for the same input and retraining FMs can result in different output for the same previously tested input \cite{10.1145/3663529.3663849}.
The simplest yet time-consuming way to evaluate FMs is to use humans knowledgeable of the ML task, while the fastest yet potentially inaccurate approach is to use the strongest available FM as a LLM-as-a-Judge. As using LLMs for evaluation does not always correlate with human evaluations \cite{liu-etal-2023-g} and introduces non-determinism into the evaluation process, we created our own metric that is deterministic and explainable.

The insight behind our metric, called \textit{Flow Similarity}, is that workflows can be represented as trees, similar to how a computer program can be represented as an abstract syntax tree. Once we have the expected and generated workflows as trees, we can use the tree edit distance algorithm \cite{doi:10.1137/0218082} to compute a similarity score. We still require humans to give us a expected workflow per user requirement, but we can automate the evaluation.

\begin{figure}[htbp]
  \centering
    \includegraphics[width=\linewidth]{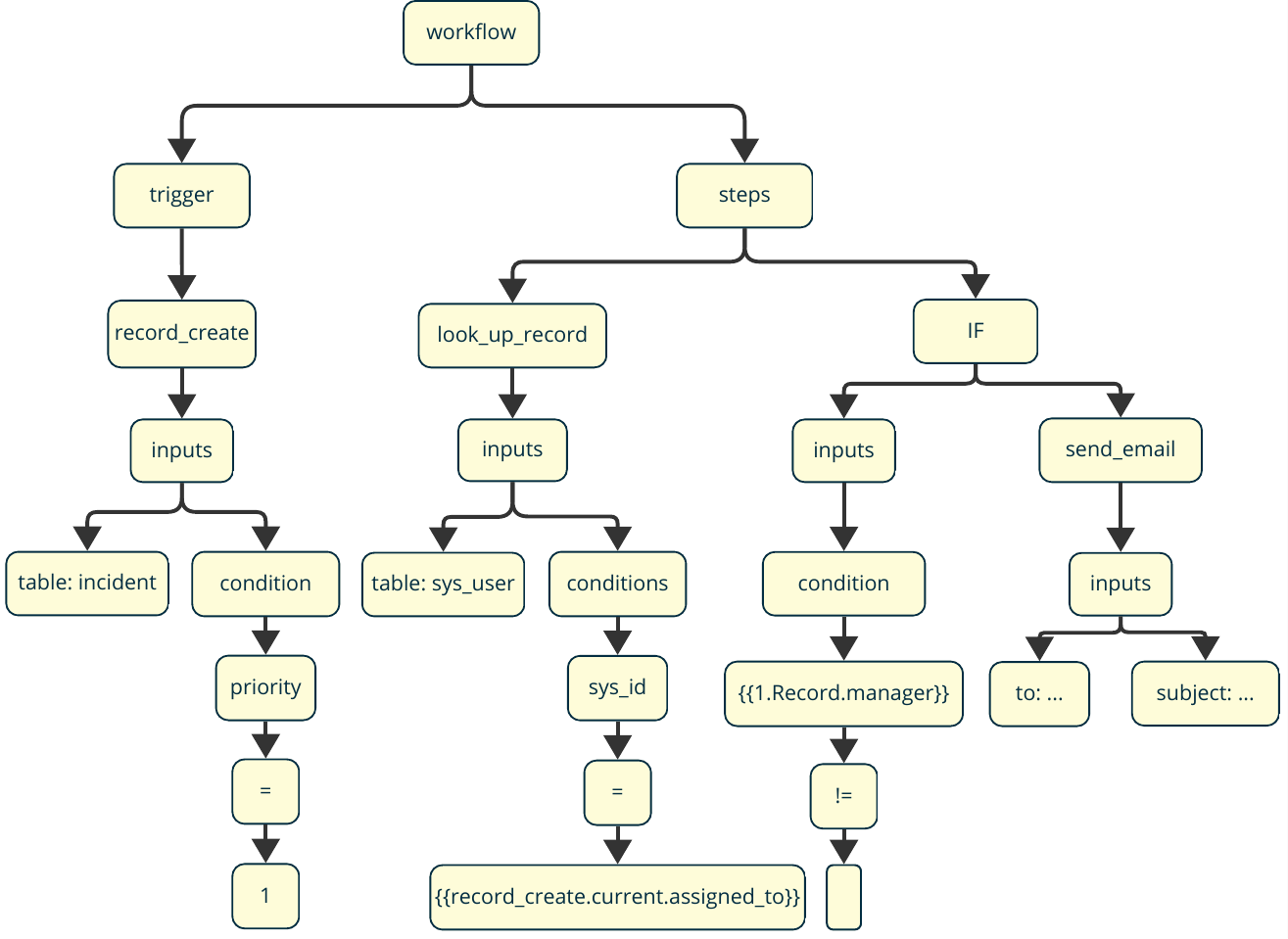}
    \caption{Tree mapping to the sample workflow in Figure \ref{fig:sample_task_example}.}
    \label{fig:flow_as_tree}
\end{figure}

Figure \ref{fig:flow_as_tree} depicts the tree representation of the workflow shown in Figure \ref{fig:sample_task_example}. Thanks to how we decomposed the tasks, we can evaluate the outline generation and the population of inputs separately, making the system highly testable. We could then decide to enable in the API the population of inputs of only the steps where we had acceptable quality.

Task Decomposition also helps the analysability of the system since we had three groups of evaluation metrics to report to our stakeholders: correctness considering only the outline, correctness considering the entire workflow, and correctness considering only step inputs (e.g., Flow similarity when populating the \texttt{look\_up\_record} step). We rank the steps by how well the FM can populate its inputs and focus on the bottom of the list.

Using RAG also offered testability benefits as we evaluated the retriever separately using recall metrics for every kind of artifact that we needed to retrieve from the environment. We analyzed the errors per kind of artifact to determine that retrieving column names is very difficult as the annotations can be of the form \textit{set it to false}. In this case, the pronoun \textit{it} refers to a previously mentioned column name. Quality can be improved by including more context in the query in addition to the annotation, such as adding the requirement.

Overall, Task Decomposition and RAG contribute to the functional correctness of the system, mainly because they allow testing and improving different parts independently: sub-tasks and the retriever. By systematically testing the different parts, we can more easily discover failure modes. This is similar to the benefits that modularity brings to building traditional software systems as engineers can devote their attention to the underperforming parts.

However, it has been previously noted that ML modularity is not the same as software engineering modularity. Improvements in a sub-task may not necessarily translate to overall system improvements as the rest of the system may not have been tuned according to the improved sub-task output \cite{8804457}. While using these two patterns enhanced the testability of the GenAI components, we still required significant end-to-end testing.

\subsection{Deployment}
The vast size of FMs in number of model parameters creates significant performance and deployment challenges \cite{10.1145/3663529.3663849}. A flexible architecture can help the deployment process by offering choices. With Task Decomposition, since each sub-task has its own level of ML difficulty and is handled separately, they can be handled by a separate FM to maximize performance.

For instance, the most complex sub-tasks can be assigned to the most powerful FM available and the simpler sub-tasks to smaller FMs, since larger models generally produce better output quality \cite{badshah2024quantifyingcapabilitiesllmsscale}. If the FMs are deployed on local servers, then the smaller FMs can be deployed on smaller, older GPUs such as V100s, reserving the more modern GPUS, A100s and H100s, for larger FMs. Or, if higher quality is needed, the most complex task can be given to a very strong cloud-served FM such as GPT-4o, while the simpler sub-tasks can be served in local servers. To further improve resource utilization, as long as the use case permits it, the sub-tasks can be batched so they can be solved concurrently, as depicted in Figure \ref{fig:td_structure_parallel}.

In our case, due to fine-tuning, we were able to obtain good quality with a small FM having only 7 billion parameters. Due to its small size, we had the same FM execute both sub-tasks on a H100 GPU. Unless computing resources are lacking, it is preferable to have the same FM execute all sub-tasks in order to reduce the deployment complexity.

Our RAG encoder is also very small having only 100 million parameters, but it achieves good retrieval quality due to fine-tuning. RAG provides good interoperability as the same encoder (and indexes) can be used for other use cases. As \textit{Workflow Generation} is not the only GenAI capability in the company's platform, the encoder was deployed as an AI component that can be used in other use cases such as Code Generation. Note that this small retriever could run on CPU due to its small size if we did not have GPUs at our disposal.

RAG however makes the deployment more complex as there are two models to deploy: the FM and the retriever encoder. If the Tool Usage pattern is used, only the FM would be deployed. However, the tools or APIs that the FM calls during generation would need to be fixed unless RAG is used to tell the FM what is available in the environment. Current industry practice shows that it is difficult to ship real-world applications with low hallucination rates and with up-to-date knowledge without a RAG implementation.

\section{Discussion}

The presented case study gives enough detail to answer RQ1 and RQ2. Task Decomposition and RAG resulted in a modular, flexible, secure, and testable system, at the expense of added complexity especially in model training. Breaking down tasks into sub-tasks requires some experimentation and AI knowledge, but it can yield faster prototyping and an iterative path to releasing GenAI applications. 

Data labeling still constitutes a significant percentage of the overall human labour in any ML project \cite{10.1007/978-3-030-64148-1_13, 10.1145/3663529.3663849}. Task Decomposition and RAG can add to the labeling effort as each sub-task, including the retrieval task, requires data at least for evaluation. However, dividing labeling tasks into smaller, simpler tasks reduces time-to-market.

Prompting an off-the-shelf FM would have resulted in faster development due to reduced labeling needs and no model training. This is preferable when the ML task is general and not domain-specific. But even when prompting, dividing the use case into sub-tasks can be beneficial. In our case, we had to fine-tune both the FM and the retriever. For more standard GenAI use cases, such as summarization or question-answering, open-source retriever models can achieve good retrieval quality. The crucial advantage with RAG, compared to other approaches that give environment data to an FM, is that it offers more control and thus security over what data can be shown to the FM.

Our model evaluation benefited from Task Decomposition and RAG as we computed evaluation metrics for each task in the GenAI pipeline: the task, the two sub-tasks, and RAG. On the other hand, these two patterns result in more deployment efforts as there are more AI components to deploy.

As our case study is limited by its self-reporting nature, we need to consider the threats to validity, considering that our objective is to show that these two techniques resulted in a well-engineered GenAI system:
\begin{itemize}
    \item \textbf{External validity:} To what extent is it possible to generalize our findings? We framed the case study by first discussing Task Decomposition and RAG in general to motivate their applicability. From our industry experience and participation in AI academic conferences, we judge that the case study will be relevant to the community and speak to current challenges. RAG is a technique that has become very common in the past year and Task Decomposition, while as a term it is not as known as RAG, is simply an application of \textit{divide-and-conquer}.
    \item \textbf{Reliability:} To what extent are the data and the analysis dependent on our own experience? To mitigate this threat, we include as much detail as possible across the main phases of the AI development cycle, including a detailed description of the task and data.
\end{itemize}

\section{Conclusion}
Motivated by our desire to bridge the AI and software engineering communities, we formalize two common techniques, Task Decomposition and Retrieval-Augmented Generation, as GenAI design patterns. They are reusable solutions to common problems and requirements arising from the need to integrate AI components into software systems. We discuss their benefits in terms of software quality attributes and consider their trade-offs and alternatives. Lastly, to illustrate their applicability, we provide a case study of building a real-world GenAI application called \textit{Workflow Generation}.

\section*{Acknowledgment}
We thank the entire team involved in shipping \textit{Workflow Generation} to end-users, hoping to have accurately distilled the lessons of the entire process in this paper.

\clearpage

\bibliographystyle{main}
\bibliography{IEEEabrv,main}

\end{document}